\documentclass[showpacs,showkeys,superscriptaddress,aps,amsmath,amssymb]{revtex4}

\usepackage{t1enc}
\usepackage{graphicx}
\usepackage{bm}

\newcommand{\cis}{$C_{IS}(t)$}
\newcommand{\cisrem}{$C_{IS}^{(rem)}(t)$}
\newcommand{\cisdif}{$C_{IS}(t) - C_{IS}^{(rem)}(t)$}
\newcommand{\tmct}{$T_{MCT}$}
    
\begin{document}
 
\title{Time correlation functions between Inherent Structures:
       a connection between landscape topology
       and the dynamics of glassy systems.}
\author{G. Fabricius} 
\email{fabriciu@fisica.unlp.edu.ar} 
\affiliation{Departamento de F\'{\i}sica, 
Universidad Nacional de La Plata,
         \\ cc 67, 1900
	La Plata, Argentina.}
\altaffiliation{Research Associate of the Abdus Salam International
Centre for Theoretical Physics, Trieste, Italy}
\author{D.A. Stariolo} 
\email{stariolo@if.ufrgs.br} 
\homepage{http://www.if.ufrgs.br/~stariolo}
\affiliation{Departamento de F\'{\i}sica, Universidade Federal 
	               do Rio Grande do Sul, \\
       CP 15051, 91501-970 Porto Alegre, Brazil}    
\altaffiliation{Research Associate of the Abdus Salam International
Centre for Theoretical Physics, Trieste, Italy}

\date{\today}

\begin{abstract}
    In this work we introduce time correlation functions between
inherent structures (IS) of a supercooled liquid.
    We show that these functions are useful to relate the 
slowing down of the dynamics to the structure of the energy landscape
near the glass transition temperature.
They show a short time regime during which the system
remains in the basin of a particular IS and a long time regime where
it explores the neighborhood of an IS. We compare the behaviour of these 
functions in a binary Lennard-Jones glass and in a model of traps
and show that they behave qualitatively different. This comparison
 reflects the
presence/absence of structure in the landscape of the Lennard-Jones/traps
models. Possible scenarios for the structure of the landscape which are
compatible with these results are discussed.
\end{abstract}

\maketitle

\section{Introduction}
 Potential Energy Landscape (PEL) studies have contributed 
 to the understanding of the dramatic slowing down of the dynamics of supercooled
 liquids near the glass transition temperature\cite{DeSt01}. 
The introduction of Inherent Structures (IS)
 (local minima of the PEL) \cite{StWe83}, which divide 
 the phase space into basins of attraction, allows the 
 separation of vibrational motion from more fundamental structural
 transitions. It is thought that activation over barriers between inherent
structures is the main mecanism of diffusion below the mode coupling transition
temperature \tmct. Above \tmct\ also saddles of the potential energy
landscape play an important role in the diffusion properties of the system.
      The decrease of the number of saddles available
  to the system when the temperature is lowered towards \tmct\ has
  been proposed as one of the main reasons of the dramatic slowing
  down of the dynamics \cite{Ca01,AnDiRuScSc00,GrCaGiPa02,AnRuSaSc03}.
In the framework of the landscape approach to the glass transition it has 
also been proposed that, at least
for some glass formers, the landscape 
  may be organized in $metabasins$ that should contain many local minima
  where the system spends increasingly long
  times as temperature is lowered \cite{St95,BuHe00}.
However, the definition of metabasins is less rigorous
  than IS and consequently even when the concept is not new
 only very recently quantitative results have begun 
  to appear \cite{DoHe03-1,DoHe03-2,DoHe03-3}.

In a previous work we studied the statistics of 
distances in real space  between neighbouring 
IS \cite{FaSt02} in a binnary mixture
Lennard-Jones glass and the number of saddles as function of temperature
following molecular dynamics trajectories. 
We saw that the distance between minima decreases exponentially near \tmct\
but does not go strictly to zero implying the existence of saddles even below
\tmct. The number of saddles suffers a strong decrease around $T=1$ and 
decays abruptly at lower temperatures but stays finite near \tmct.
These results together with the vanishing of the diffusivity at \tmct\ 
seem to imply the existence of saddles associated with nondiffusional events.
  The picture that emerges is one where the system spends
  long times in a kind of metabasins of the energy landscape
  each one of them containing many IS similar to the one proposed by
Doliwa and Heuer \cite{DoHe03-1,DoHe03-2,DoHe03-3}.

Here we show that some dynamical quantities in two systems which have 
a similar kind of glass transition but posses very different PEL structure 
behave in a qualitatively different way. This allows to distinguish the
mechanisms of confinement in both kind of systems. We define time
correlation functions between inherent structures visited during the
equilibrium time evolution of a binary mixture Lennard-Jones glass. These
correlations present two sharply distinguished time regimes which we show
to correspond to exploration inside a basin of a particular IS and to
exploration of the neighborhood of a basin. The behaviour of the correlations
as temperature is lowered towards \tmct\ implies the emergence of strong
spatial confinement in configuration space. In order to test if these
confinement is a consequence of the topology of the PEL or if it can be
explained as a purely dynamic effect we compare the behaviour of the
Lennard-Jones glass with a model of traps where the landscape is flat and
only confinement in time is present. In fact both systems behave qualitatively
different implying that confinement in the Lennard-Jones glass is a consequence
of the non trivial structure of the energy landscape. We discuss some 
possible scenarios for this structure which are compatible with the dynamical
characteristics observed. While the existence of metabasins is a plausible
scenario it is not the only possibility.

In section \ref{correlations} we define the quantities of interest in our
study. In section  \ref{LJ}
we show and discuss our results for the binary Lennard-Jones model. In
section \ref{traps} we show the corresponding results for a model of traps
and finally in section \ref{final} we present our conclusions.

\section{Time correlation functions between Inherent Structures}
\label{correlations}
As mentioned above every instantaneous state of the system
may be associated with an IS of the energy
landscape.
Given that at an initial time the system is in a configuration corresponding to some
IS, we define $C_{IS}(t)$ as the probability that the system 
{\em is at the same} IS after a time $t$.
This means that the IS associated with the inital and final states is
the same irrespective of where the system was in between.
We have observed that the IS
can be identified by their energy
since the case of degeneracy (interchange of two particles of the same kind,
for example) is quite unusual and has no significant statistical weight.
So, the computation of $C_{IS}(t)$ is very simple:
\begin{equation}
C_{IS}(t) =\frac{1}{N_t} \sum _{i=1}^{N_t} \delta _{t_i , t_i + t}
~ , ~~~~~{\rm with} ~~ \delta _{t_i , t_j}=
           \left\{
	   \begin{array}{l}
	   1 ~ {\rm if~ } E_{IS}(t_i)=E_{IS}(t_j)\\
	   0 ~ {\rm if~  not} 
	   \end{array} \right.
\label{eqcis}
\end{equation}
where $E_{IS}(t_i)$ is the energy of the corresponding IS at time $t_i$
and with the sum we perform an average over $N_t$ available times.
It will be useful to study also another time correlation function,
$C_{IS}^{(rem)}(t)$,
given by the probability that the system
{\em has remained} at the same inherent structure during a time $t$.
In this case $\delta _{t_i , t_j}$=1 only if $E_{IS}(t_i)=E_{IS}(t_k)$
for all $t_k \in (t_i,t_j)$.
In fact, the calculation of $C_{IS}^{(rem)}(t)$ involves an approximation
due to the discretization of time. If the system goes to a neighbouring
IS and comes back within the interval $(t_i,t_{i+1})$ our calculation 
won't notice it, so the calculated C$_{IS}^{(rem)}$(t) should be taken
as an upper bound of the real one. Other kinds of correlation functions
between inherent structures have been already explored in the literature
\cite{CrRi00,BeGa03}, although we prefer this new one which allows to
illustrate clearly the connection with the topology of the landscape, as
will be shown below.

\section{The Lennard-Jones binary mixture}
\label{LJ}
\subsection{Size dependence of $C_{IS}(t)$}
Unfortunately these time correlation functions are size dependent.
Consider a system $S_0$ of $N$ particles compossed by two non-interacting
subsystems $S_1$ and $S_2$ of $N_1$ and $N_2$ particles respectively.
In this case $C_{IS}(t,N)=C_{IS}(t,N_1)\cdot C_{IS}(t,N_2)$,
since the probability of coming back to a given state $IS_0$ involves
the two subsystems coming back to their respective states $IS_1$
and $IS_2$.
In the present work we study a system of $N=130$ particles
at several temperatures. In order to study the size dependence
of $C_{IS}(t)$ we performed also a simulation at $T=0.55$ 
for a system with $N=250$ particles.
\begin{figure}[ht]
\includegraphics[width=5.5cm,height=6.8cm,angle=270]{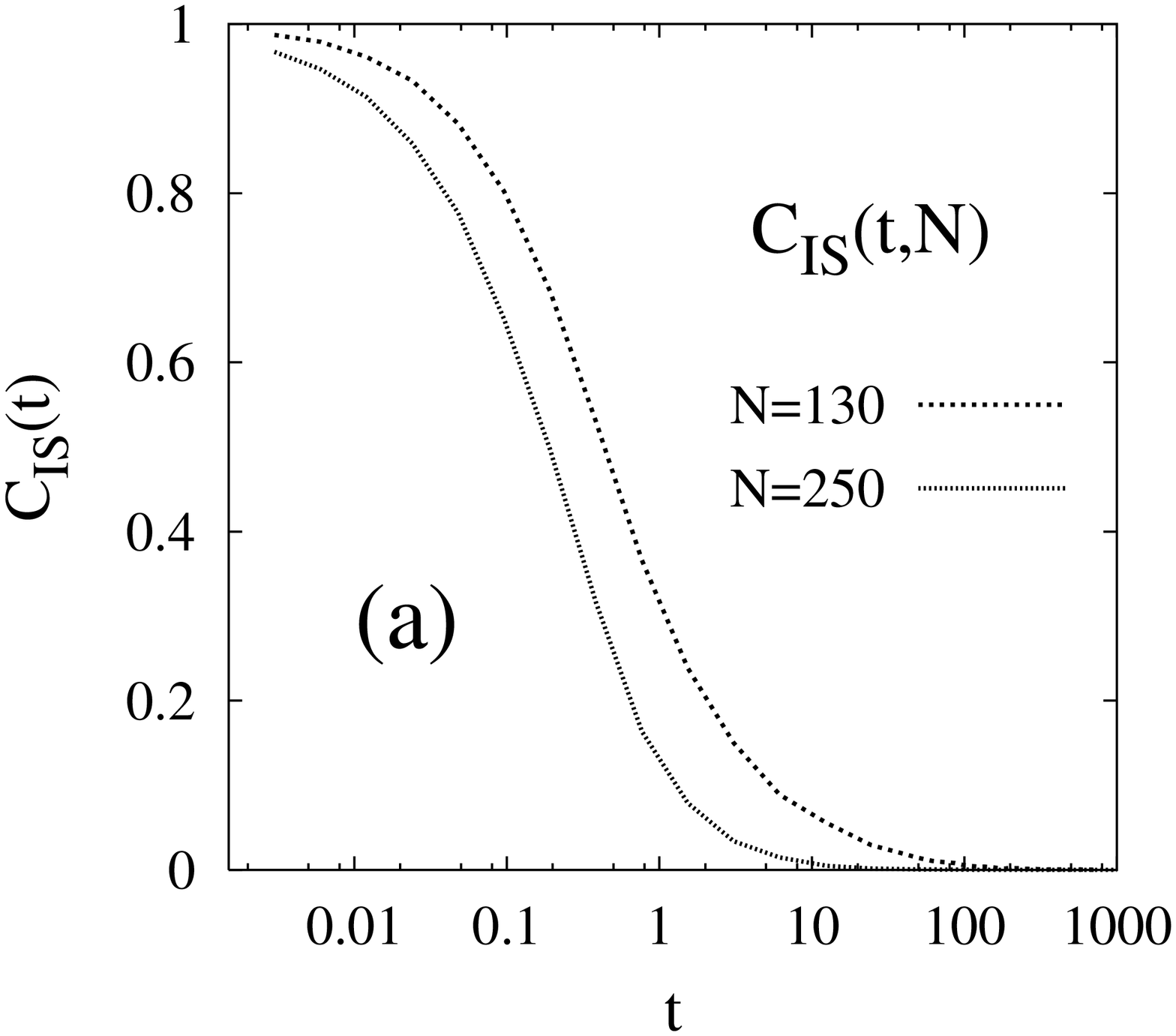}
\includegraphics[width=5.5cm,height=6.8cm,angle=270]{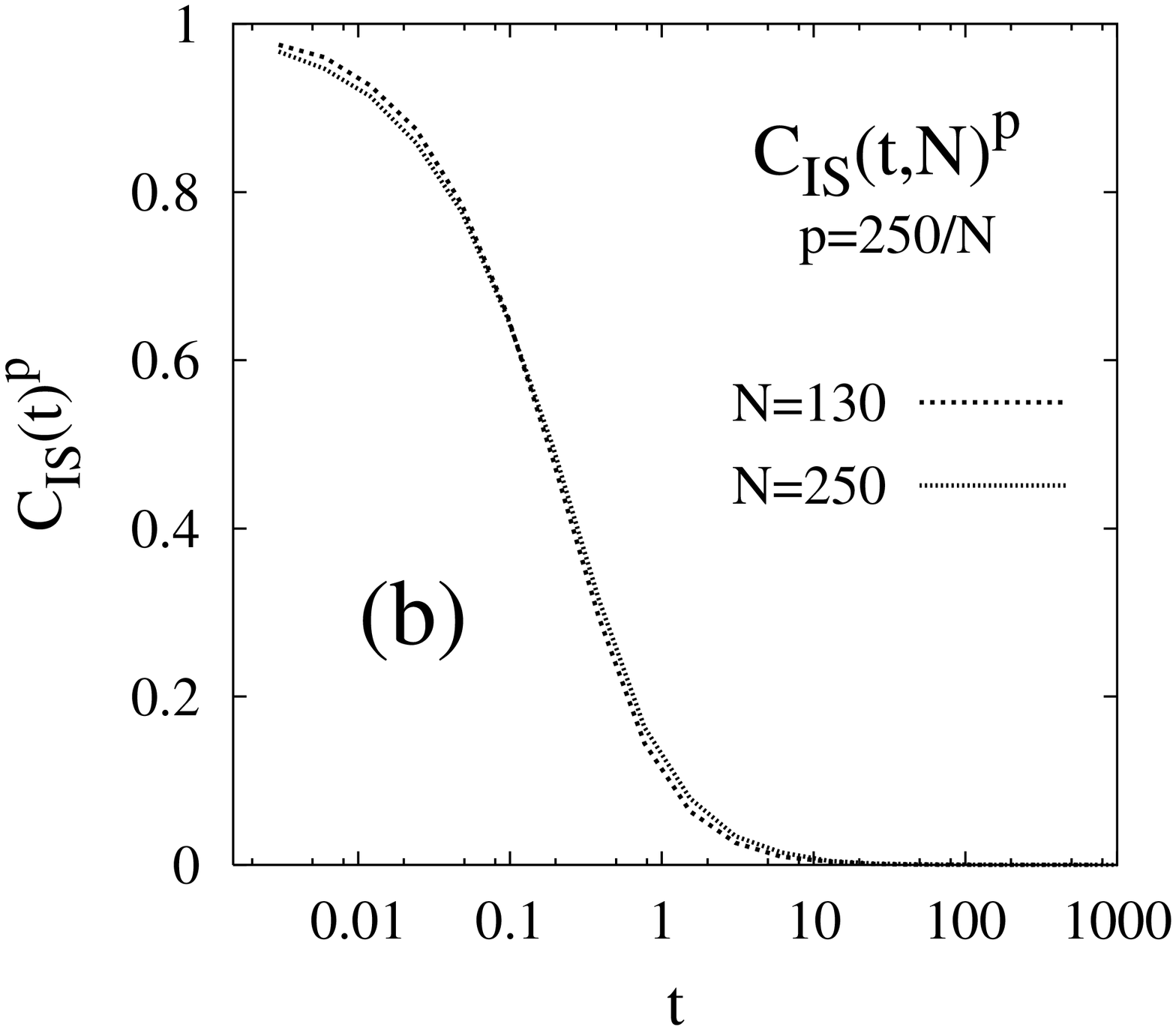}
\caption{\label{size} 
Size dependence of $C_{IS}(t)$ at $T=0.55$.
}
\end{figure}
In figure 1a we show $C_{IS}(t,N)$ for the two sizes
considered. If the hypotesis of independence between subsystems
holds for subsystems of size $N>N_0$ then it is easy to
show that for $N >N_1 > N_0$: 
\begin{equation}      
C_{IS}(t,N)=\left[C_{IS}(t,N_1)\right]^{\frac{N}{N_1}}
\end{equation}      
In figure
1b we show that this equality is verified quite well for the
systems studied. Studying small systems is convinient because
the decay of $C_{IS}(t,N)$ is slower for smaller N, an so one may
access the long time behaviour of this correlation function. This
fact has also been emphasized in other landscape studies
of supercooled liquids \cite{DoHe02,DoWa02}.
Having shown that extrapolation 
to large systems is quite simple, 
in what follows we will study the properties of
correlation functions for the $N=130$ system. It is important to
note that, because of the size dependence,  the relaxation times of 
these correlation functions 
can not be straightforwardly related to the typical time scales of the 
bulk relaxations in supercooled liquids.

\subsection{Temperature dependence of C$_{IS}$(t)}
\label{results}
Here we show results of molecular dynamics simulations of a binary mixture
(80:20) Lennard-Jones system \cite{FaSt02} for $N=130$
at temperatures: $2.0, 1.0, 0.8, 0.7, 0.6, 0.55$ and $0.5$. We have
verified that the mode coupling glass transition for this system is the 
same as found originally for a sample of 1000 particles ($T_{MCT}=0.435$ 
\cite{KoAn95}).
We repeat the study for two samples with different
thermal histories.
After having equilibrated the systems we performed long runs 
in order to collect data \footnote{Tipically t$_{run}\sim$ 1000 $\tau$. 
The MD time step in our
simulations is 0.003 Lennard-Jones time units which amounts to runs of $2.10^7$ 
time steps for the smaller temperature studied .}.
, 
we obtained
the IS relaxing the instantaneous coordinates of the system
and we computed
the time correlation functions from the sequence of $E_{IS}(t_i)$
(equation 1).
\begin{figure}[ht]
\includegraphics[width=5.5cm,height=6.8cm,angle=270]{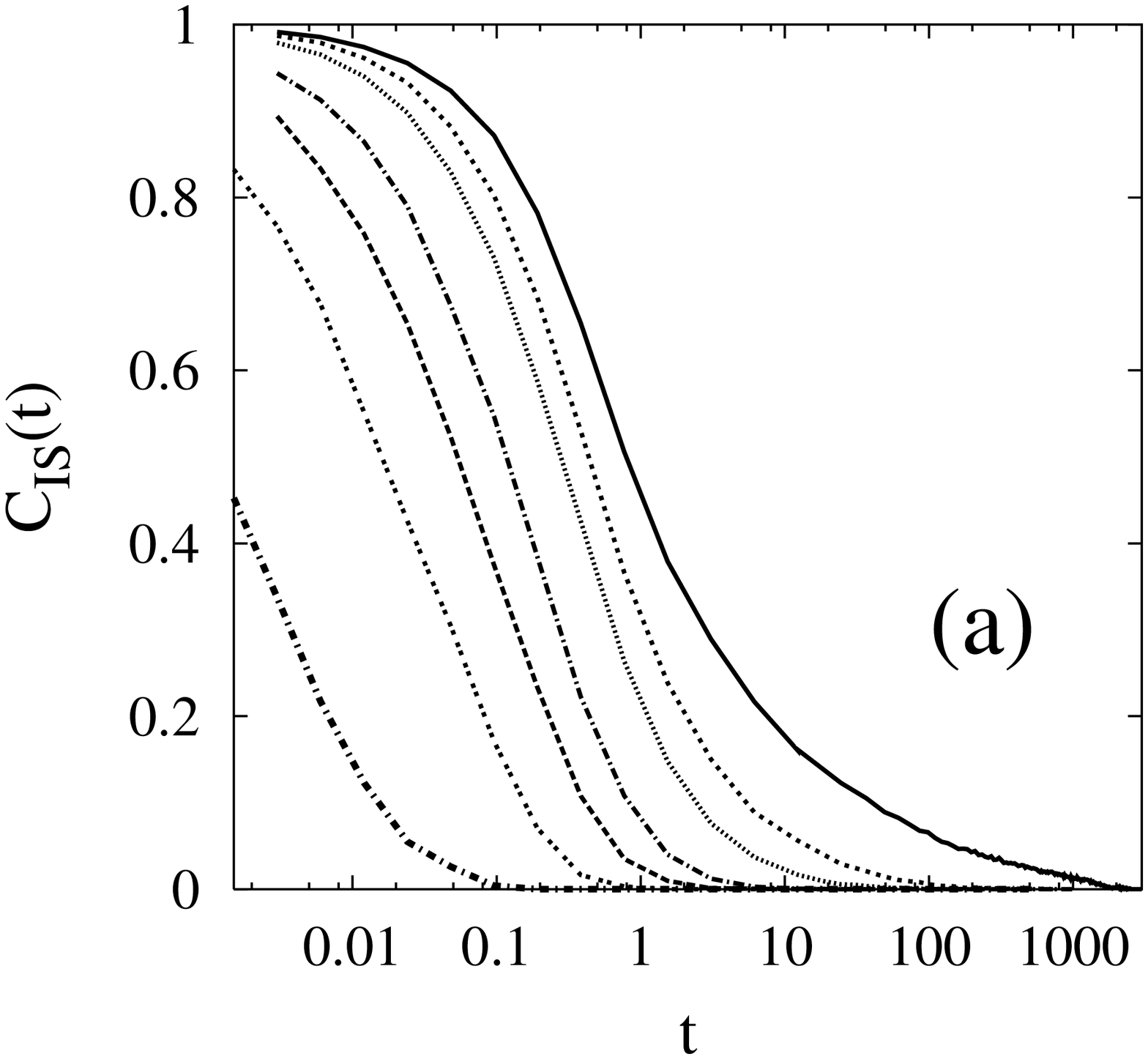}
\includegraphics[width=5.5cm,height=6.8cm,angle=270]{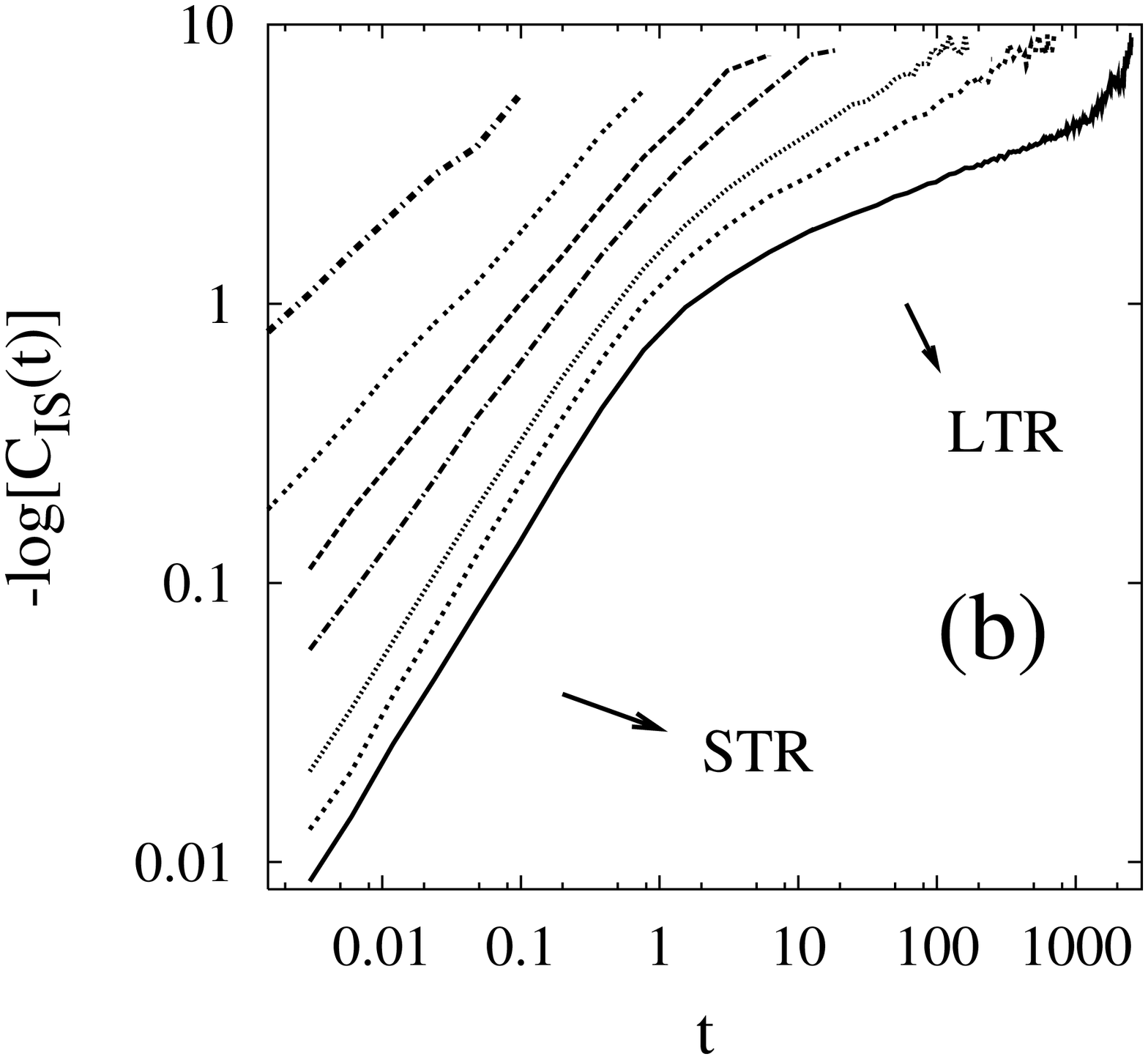}
\caption{\label{temp} 
(a) Temperature dependence of $C_{IS}(t)$ for $N=130$.
From left to right $T=2.0, 1.0, 0.8, 0.7, 0.6, 0.55$ and $0.5$.
(b) -log [ $C_{IS}(t)$ ] for the same set of temperatures; 
plotted in this way a stretched exponential
$exp[ -(t/\tau)^\beta$ ] is a straight line with slope $\beta$. 
The short time regime (STR) and  
long time regimes (LTR) are indicated.
}
\end{figure}

In figure 2a we plot $C_{IS}(t)$ for the set of temperatures
considered. It is clear that
the decay of $C_{IS}(t)$ slows down very quickly as $T_{MCT}$
is approached. Figure 2b shows that at low temperatures
$C_{IS}(t)$ may be 
approached by two stretched exponentials with quite 
different exponents. In the short time regime (STR) the value of $\beta$
is around $0.8$ and it is almost constant for $T$ varying between 0.6 and 0.5.
In the long time regime (LTR) the value of beta is very small ($\beta=0.2$
 for $T=0.5$) and is decreasing sharply on approaching  $T_{MCT}$.
The deviation from stretched exponential behaviour at the longer times of 
figure 2b for $T=0.5$ is due to limited statistics. We have verified in some
longer runs that the stretched exponential behaviour extends at least for the
whole range of times of figure 2b.
\begin{figure}[ht]
\includegraphics[width=5.5cm,height=7.8cm,angle=270]{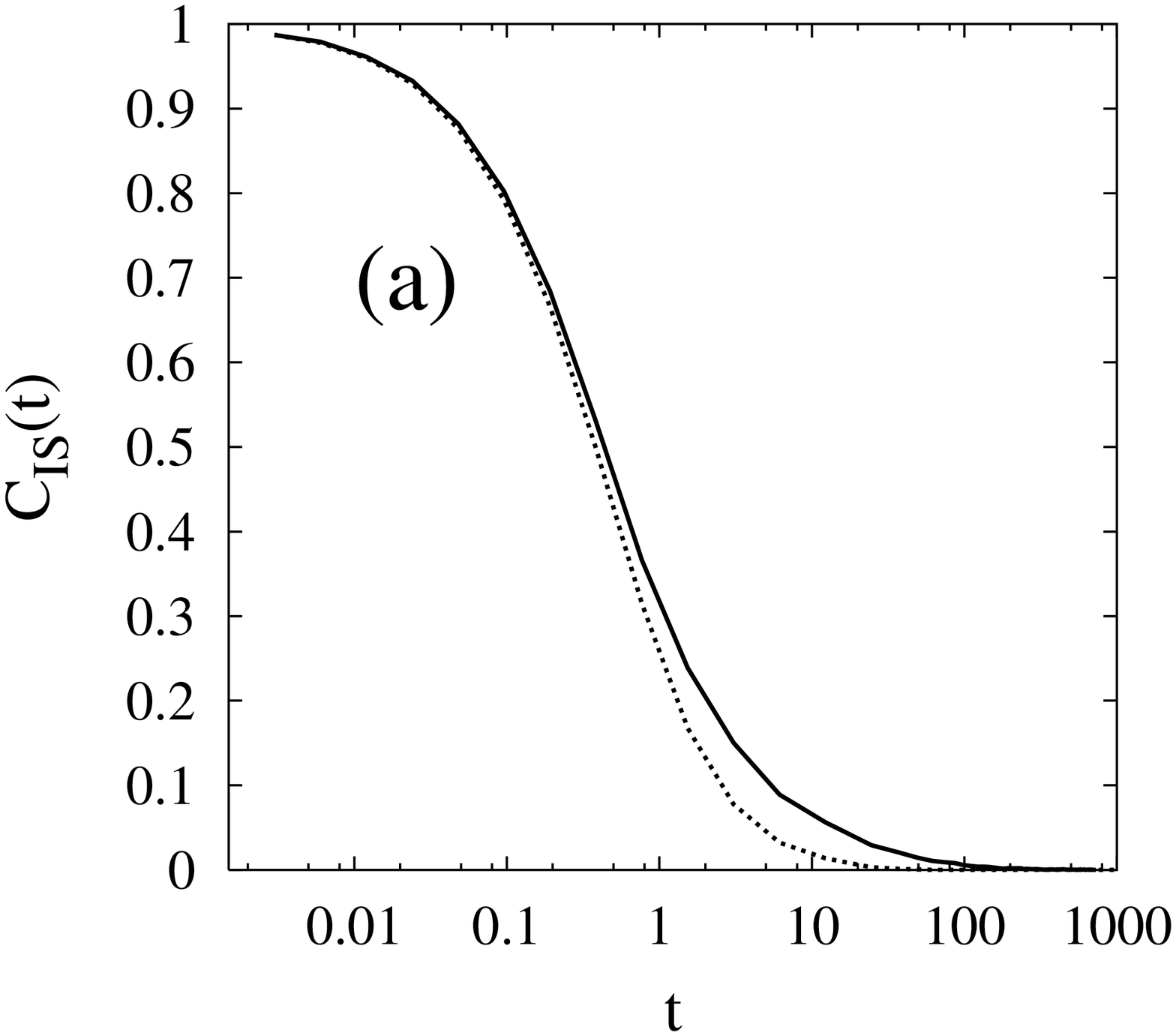}
\includegraphics[width=5.5cm,height=6.2cm,angle=270]{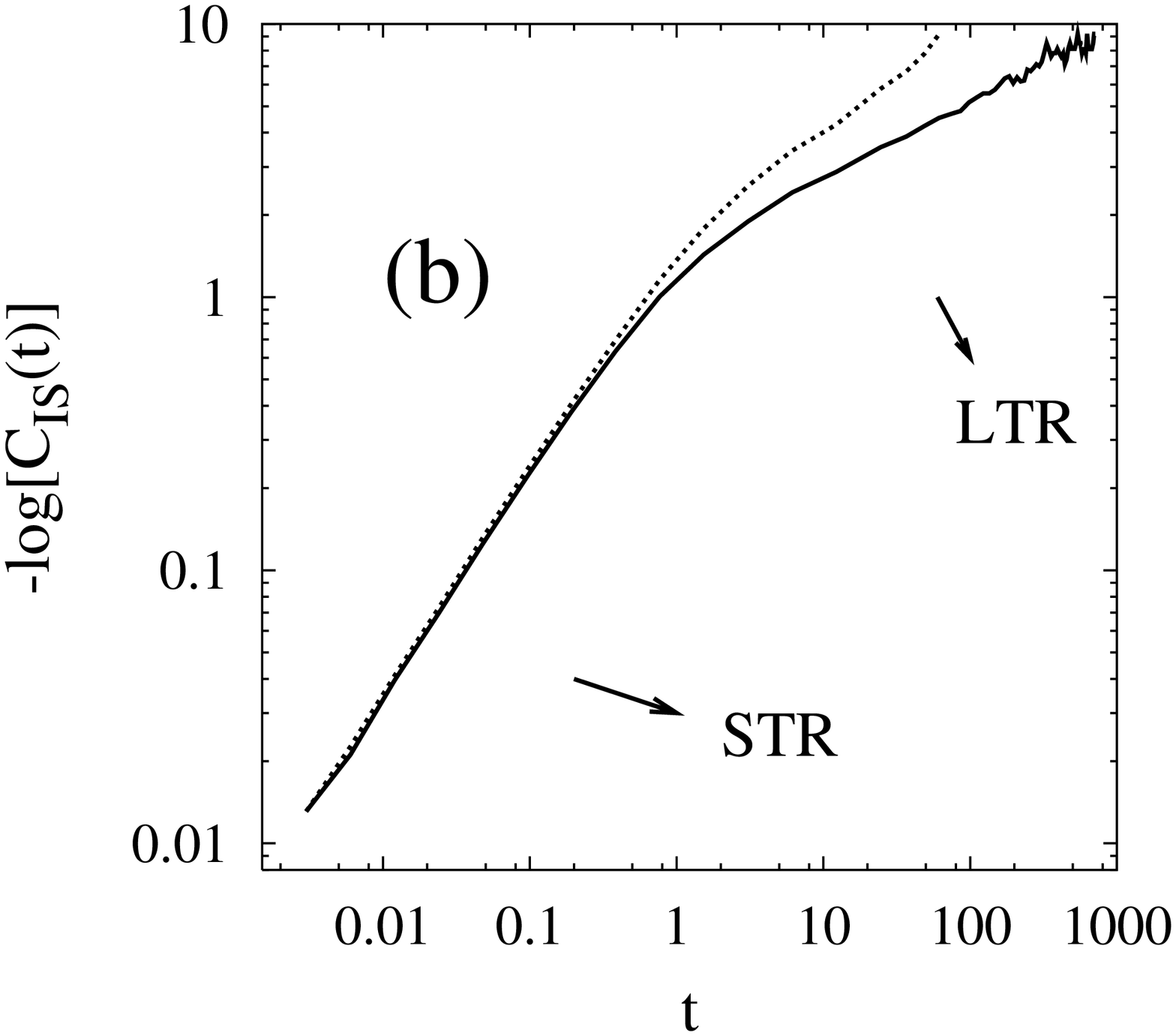}
\caption{\label{rem} 
$C_{IS}(t)$ (continuous line) vs $C_{IS}^{(rem)}(t)$ (dotted line) 
for $N=130$ and $T=0.55$.
The fit to stretched exponentials of $C_{IS}(t)$ gives
$\beta=0.83$ for the STR and $\beta=0.28$ for the LTR.
The inset of (a) shows the difference:
$C_{IS}$ - $C_{IS}^{(rem)}$ for $T=0.7, 0.6, 0.55$ and $0.5$ 
(from left to right). 
}
\end{figure}
In order to explain the meaning of the two regimes observed we show
in figure 3
$C_{IS}(t)$ together with $C_{IS}^{(rem)}(t)$ for $T=0.55$.
We see that the STR is the region of time
where both correlation functions are almost equal. So
the STR of $C_{IS}(t)$ may be interpreted as the
time the system remains at the same IS while the LTR
accounts for the time the system is $around$ one IS,
exploring neighbouring inherent structures  
and with a finite probability of coming back to the original IS.

\begin{figure}[ht]
\includegraphics[width=5.5cm,height=7.8cm,angle=270]{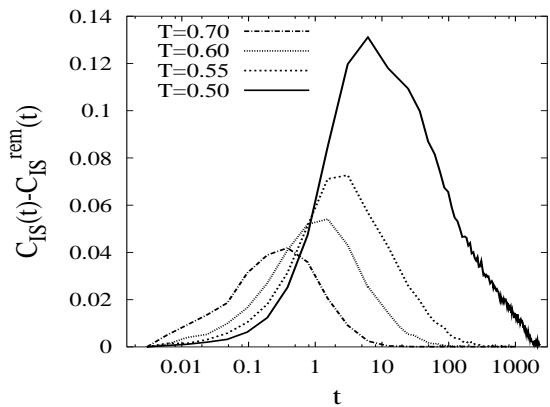}
\caption{\label{dif} The difference
$C_{IS}$ - $C_{IS}^{(rem)}$ for the LJ binary mixture for 
different temperatures. }
\end{figure}

In order to filter from the $C_{IS}(t)$ the effect of remaining in a single
basin all the time, we show in  figure \ref{dif} the difference
$C_{IS}$ - $C_{IS}^{(rem)}$ which represents the probability of the system to be at 
the same IS after a time $t$ {\em knowing that it has gone away 
at least once during this time interval}. 
This probability shows two important characteristics: the first one is
that there is a characteristic time at which the probability of coming back
to the original IS is maximal. This time scale grows
when temperature is lowered: if the system goes out of a basin it takes more time
to come back as temperature is lowered. The second feature is more
important: the probability to come back {\bf grows} as the glass transition
temperature is approached. This is a signature of {\em spatial confinement}.
To show more clearly this point and closely inspect the relation between the
studied correlation functions and landscape topology we study in
the next section a random walker on a lattice with traps.

\section{The Traps Model}
\label{traps}
The model of traps in a d-dimensional hypercubic lattice is realized as a
random walk of a particle hopping between traps attached at each lattice
site with a given trap energy distribution $\rho(E)$ \cite{MoBo96,BoGe90}. 
Depending on the form of
the distribution different interesting dynamical behaviours are observed.
If the energies are exponentially distributed the model has a dynamical
phase transition at a finite temperature $T_0$ below which it presents
typical glass phenomenology as aging effects. Other possibilities like
e.g. gaussian distribution of energies lead to a dynamical transition only
at $T_0=0$ and show stretched exponential relaxations. 
The trap model has been proposed as a
phenomenological or toy model for describing the physics of glasses
and spin glasses. From a physical point of view the dynamics proceeds 
through activation over barriers corresponding to the depth of the traps.
Besides the difference in the depths the landscape can be considered flat,
structureless. This is the essential point we want to compare with the
Lennard-Jones glass which is known to possess a complex PEL topology. In 
an interesting study Denny et al.\cite{DeReBo02} compared some properties
of the Lennard-Jones glass with the model of traps.  They found that there
are collections of IS of nearly the same energy which they associated with 
metabasins and
present a nearly gaussian distribution of energies, similar to the distribution
of single inherent structures energies. Some relaxation functions do indeed
behave in agreement with results on a model of traps, a result which implies
that activated processes are important in the dynamics of the LJ binary mixture
also above the mode coupling transition temperature.

Here we also make a comparison between the Lennard-Jones glass and the 
model of traps. But our aim here is to calculate the time correlation
functions defined in section \ref{correlations} for a simple model
where the probability of the system to return to a given basin
is exclusively of random character. So, we associate a trap to a basin 
(not to a collection of basins like
Denny et al. did) and study the probability that the system returns
to a given basin.
We considered  two and three dimensional lattices, and assign to 
every site a trap of energy E that is determined randomly
from an exponential distribution of the form $e^{E/T_0}$. 
The energy associated to a given
site is kept fixed during the simmulation, i.e.
when the walker returns to a given site it finds
the same trap (quenched-disorder case\cite{BoGe90}).
In order to make the comparison with the results for
the binary LJ model as closer as possible we choose the typical energy of the
exponential distribution as $T_0$=\tmct\ , so both models
are expected to have a dynamical transition at the same temperature
of \tmct\ =0.435. Because of computational limitations periodic boundary 
conditions in a lattice of size $L^d$ are imposed to the values of the energies
 although the walker
moves in an infinite lattice limited in practice only by the largest time
considered in the correlation functions.
We have used $L=100$ for $d=3$
and $L=1000$ for $d=2$. To obtain enough statistics at 
the lowest temperatures studied for $d=2$ we needed to perform
rather long walks of $10^8$ steps that we averaged over
1000 realizations of disorder. We then computed the time correlation
functions defined in \ref{correlations} along these runs. In fact,
because of the periodic boundary conditions imposed in the energies,
in this case we use the position of a site instead of its energy 
in order to compute de delta function in expression \ref{eqcis}.
Note that the function $C_{IS}^{(rem)}(t)$ defined above corresponds exactly
to the equilibrium correlation $C_{eq}(t)$ defined in equation (4) of
\cite{MoBo96}.
We verified that \cisrem\ presents the expected long time behaviour
at low temperatures: $C_{IS}^{(rem)} \propto t^{-(x-1)}$ with
$x=T$/\tmct\ , obtained theoretically by Monthus {\it et al}
\cite{MoBo96} (equation (14)).

\begin{figure}[ht]
\includegraphics[width=5.5cm,height=7.8cm,angle=270]{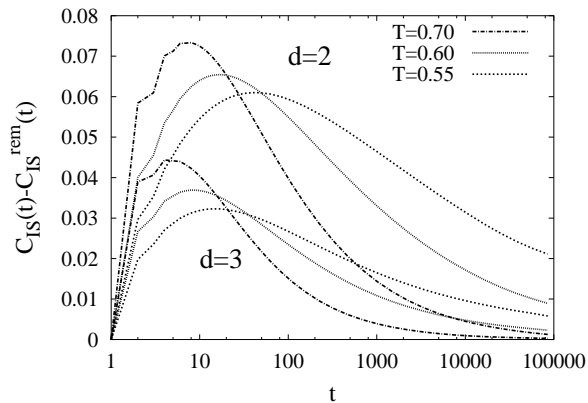}
\caption{\label{trap} The difference
$C_{IS}$ - $C_{IS}^{(rem)}$ for the traps model in  
$d=2$ and $d=3$ and the same temperatures of figure \ref{dif}. }
\end{figure}

In figure \ref{trap} we show the difference $C_{IS}$ - $C_{IS}^{(rem)}$ 
for the traps model in two and three dimensions.
The behaviour of these correlation functions
as a function of temperature is
qualitatively the same for d=2 and d=3. They present 
a characteristic time that increases as temperature is lowered
as we observed for 
the curves of figure 4 corresponding to the binary LJ model.
In the case of traps, the maximum of the probability of
coming back to a given trap moves towards increasing times as 
temperature is lowered because the walker is caught increasing times
in the traps of the sorroundings. 
The confinement can only be attributed to the time 
the walker spends
in the individual traps since every trap is spatially 
equivalent to each other, there is no spatial confinement
as the walker moves in a flat landscape.
More importantly, the fact that the height of the maximum of these curves
increases for the binary LJ model while it goes down
for the traps-model is a clear indication that an adittional
type of
confinement exists in the energy landscape of the LJ model.
An arrangement of the energy landscape in metabasins 
that contain several basins 
(as has been suggested by Doliwa {\it et al.} \cite{DoHe03-1,DoHe03-2,DoHe03-3})
could explain the qualitative features of the curves
 we observe in figure 4 since the system would be forced
 to explore several IS within a metabasin before leaving it
 for a time that increases on approaching $T_{MCT}$.
 But inspection  of figure 5 suggest another possible
 picture for the topology of the energy landscape.
 From figure 5 is clear that reduction of the dimention
of the space where the walker moves increases 
sharply the absolute value of the studied probabilities.
So, another possible mechanism that may account for 
the increasing confinement of the binary LJ system is 
a reduction of dimensionality of the 
 space where the system moves. In this
 picture, as temperature lowers, the system is forced 
 to move in a landscape of reduced dimensionality
 where the different accesible regions of configuration
 space need not be sepparated by any energy barrier.

\section{Summary and conclusions}
\label{final}

In the present paper we defined  time correlation functions
between inherent structures of an energy landscape and study them 
for a LJ binary mixture at different temperatures.
We observed that the probability of being at the same
IS after a time $t$, \cis , presents two time regimes:
a short time regime which is related to exploration of basins,
and a long time regime which corresponds to exploration of the neighborhood
of a given basin. 
The weak dependence of $\beta$ on temperature
for the  STR 
at low temperatures indicates that the basin topology
doesn't change on approaching $T_{MCT}$, while the contrary
holds for the inter-basins topology given the strong variation of
$\beta$ with $T$ in the LTR.
The study of the difference \cisdif\ for
this system and comparison of the same function
for a traps-model give clear evidence
that in the LJ model there is a spatial mechanism of confinement
that is not present in the traps model. 
From our results two possible mechanisms emerge to account
for the confinement at low temperatures. One possiblity is
that inherent structures are arranged in superestructures like
in the metabasins scenario\cite{DoHe03-1,DoHe03-2,DoHe03-3}.
This scenario implies as a basic ingredient the existence of
energy barriers, necessary to define the metabasins and consequently
the landscape appears to be broken in pieces (metabasins). 
Nevertheless the confinement observed in our results can also
be explained by a gradual reduction of the dimensionality 
accesible to the system as the temperature is lowered. In this
second scenario, as energy barriers are not implied a priori,
the landscape is not necessarily broken in metabasins but its
structure can be that of a network composed of connected funnels. 
In this scenario the confinement is consequence of a reduction
of the dimension of the funnels as temperature is lowered.

We have shown that the time correlation functions introduced in
this paper are
useful tools to get information on the structure of the PEL in supercooled
liquids and can also be straightforwardly calculated in trap
like models. It would be interesting to analyze their behaviour
in other models in which one could a priori define some structure
or hierarquical organization of states which seems to be present
in the LJ binnary mixture.

This work was supported in part by CONICET and {\it Fundaci\'on Antorchas}, Argentina
and by CNPq, Brazil.


\end{document}